# Связь параметров критической точки фазового перехода газ-жидкость с температурой Бойля

Умирзаков И. Х.

*Институт теплофизики, Новосибирск, Россия*
*e-mail:* [tepliza@academ.org](tepliza@academ.org)

**Аннотация**

Показано, что для 21 реального вещества отношение температуры Бойля к произведению критической температуры и критического фактора сжимаемости с высокой точностью равно числу 9, предсказываемому уравнением состояния Ван-дер-Ваальса. Предложено соотношение для нахождения критического объема через отношение температуры Бойля к критическому давлению. Предложена также формула для определения критического объема через критическую температуру и параметры прямой линии единичной сжимаемости.

***Ключевые слова****: критическая температура, критический объем, критическое давление, температура Бойля, уравнение состояния Ван-дер-Ваальса, фазовый переход жидкость-газ.*

## Введение

На основе своего знаменитого уравнения состояния $p(T,v) = kT/(v-b) - a/v^2$, (1)

где $k$ – постоянная Больцмана, $T$ – температура, $v$ – объем, приходящийся на одну частицу (молекулу или атом), $p$ – давление, $a$ и $b$ – постоянные положительные параметры,

Ван-дер-Ваальс впервые объяснил существование критической точки фазового перехода жидкость-газ [1].

На основе уравнения Ван-дер-Ваальса (1) теоретически было получено уравнение линии единичной сжимаемости (на которой выполняется уравнение состояния идеального газа) в виде линейной связи между температурой и плотностью [2]. Эта связь имеет место для многих полярных и неполярных веществ различного химического состава. В работе [3] на основе этого уравнения был теоретически выведен эмпирический закон Кальете-Матиаса (или правило «прямолинейного диаметра»). Поэтому можно ожидать, что уравнение Ван-дер-Ваальса может дать новые связи между термодинамическими характеристиками веществ и материалов. Настоящая работа посвящена нахождению таких новых связей.

Поиск связей между термодинамическими характеристиками веществ является целью термодинамической теории подобия [4]. Эти связи являются основой прогнозирования теплофизических свойств веществ [5]. Актуальность поиска таких связей обусловлено увеличением количества используемых веществ и материалов, дороговизной и трудоемкостью изучения их свойств, а иногда и невозможностью изучения их в интересующей области значений термодинамических параметров.

## Результаты и их обсуждение

### 1. Связь критического фактора сжимаемости с отношением температуры Бойля к критической температуре для газа Ван-дер-Ваальса

Из уравнения состояния Ван-дер-Ваальса (1) следует, что:
- второй вириальный коэффициент $B(T) = b - a/kT$;
- температура Бойля $T_B$, определяемая уравнением $B(T_B) = 0$ (стр. 19 в [6]), равна

$$T_{Bvdw} = a/kb;$$ (2a)

- критическая температура $T_c$ и критический объем в расчете на одну частицу $v_c$, полученные из условий термодинамической устойчивости в критической точке $\partial p(T,v)/\partial v|_{T_c,v_c} = 0$ и $\partial^2 p(T,v)/\partial v^2|_{T_c,v_c} = 0$ (стр. 35 в [7]), равны $v_{cvdw} = 3b$, $T_{cvdw} = 8a/27kb$; \hfill (2b)

- критическое давление $p_c = p(T_c,v_c)$ и критический фактор сжимаемости $z_c = p_c v_c / kT_c$ равны (стр. 211 в [8]) равны $p_{cvdw} = a/27b^2$, $z_{cvdw} = 3/8$. \hfill (2c)

Здесь $vdw$ в индексах величин означает, что соответствующая величина относится к уравнению Ван-дер-Ваальса.

Из равенств (2a)-(2c) имеем $T_{cvdw}/T_{Bvdw} = 8/27 = 0.296$, $1/z_{cvdw} = 8/3 = 2.667$, \hfill (3)

$(T_{Bvdw}/T_{cvdw})/z_{cvdw} = 9$. \hfill (4)

Исходя из (4) можно предположить, что для реальных веществ тоже выполняется равенство $(T_B/T_c)/z_c = 9$. \hfill (5)

## 2. Связь критического фактора сжимаемости с отношением температуры Бойля к критической температуре для реальных веществ

Чтобы проверить выполняемость соотношения (5) для реальных веществ, мы для 21 реального вещества из табл. 1 провели статистический анализ, результаты которой представлены в табл. 2. В табл. 1 значения критической температуры и критического фактора сжимаемости взяты из [7], а значения температуры Бойля – из [9]. В табл. 2 введены обозначения: $<f> = \sum_{j=1}^{N} f_j / N$ – для среднего значения величины $f$;

$\delta f = \sqrt{\sum_{j=1}^{N} (f_j - <f>)^2 / N}$ – для дисперсии – среднеквадратичных отклонений $f$ от её среднего значения;

$\Delta f = \sum_{j=1}^{N} |f_j - <f>| / N = \sum_{j=1}^{N} \sqrt{(f_j - <f>)^2} / N$ – для среднего абсолютного значения отклонений $f$ от её среднего значения, $N = 21$.

Из табл. 2 легко получить, что $<(T_B/T_c)/z_c> = 8.978$, $<(T_B/T_c)>/<z_c> = 8.982$ и $<(T_B/T_c)> \cdot <1/z_c> = 8.993$. Они отличаются от числа 9 на 0.2%, 0.2% и 0.1%, соответственно.

**Табл. 1.** Температура Бойля $T_B$, критическая температура $T_c$, критический фактор сжимаемости $z_c$, отношений $T_B/T_c$, $1/z_c$ и $T_B/T_c/z_c$ для 21 вещества различного химического состава

| Вещество | Символ | $T_B, K$ | $T_c, K$ | $z_c$ | $T_B/T_c$ | $1/z_c$ | $T_B/T_c/z_c$ |
|---|---|---|---|---|---|---|---|
| Метан | $CH_4$ | 509.74 | 190.56 | 0.288 | 2.675 | 3.472 | 9.288 |
| Этан | $C_2H_6$ | 768.49 | 305.33 | 0.285 | 2.517 | 3.509 | 8.831 |
| Этилен | $C_2H_4$ | 724.33 | 282.35 | 0.276 | 2.565 | 3.623 | 9.295 |
| Пропан | $C_3H_8$ | 902.56 | 369.83 | 0.281 | 2.440 | 3.559 | 8.685 |
| Бутан | $C_4H_{10}$ | 1017.7 | 425.13 | 0.274 | 2.394 | 3.650 | 8.737 |
| Изобутан | $C_4H_{10}$ | 984.54 | 407.82 | 0.283 | 2.414 | 3.534 | 8.531 |
| Пентан | $C_5H_{12}$ | 1113.5 | 469.7 | 0.262 | 2.371 | 3.817 | 9.048 |
| Аргон | $Ar$ | 407.76 | 150.69 | 0.291 | 2.706 | 3.436 | 9.299 |
| Окись углерода | $CO$ | 341.32 | 132.9 | 0.295 | 2.568 | 3.390 | 8.706 |
| Двуокись углерода | $CO_2$ | 717.93 | 304.13 | 0.274 | 2.361 | 3.650 | 8.615 |
| Дейтерий | $D_2$ | 113.7 | 38.34 | 0.314 | 2.966 | 3.185 | 9.444 |
| Криптон | $Kr$ | 569.61 | 209.48 | 0.288 | 2.719 | 3.472 | 9.442 |
| Неон | $Ne$ | 120.32 | 44.49 | 0.311 | 2.704 | 3.215 | 8.696 |
| Азот | $N_2$ | 326.41 | 126.19 | 0.29 | 2.587 | 3.448 | 8.920 |
| Закись азота | $N_2O$ | 754.8 | 309.58 | 0.274 | 2.438 | 3.650 | 8.898 |
| Кислород | $O_2$ | 408.35 | 154.58 | 0.288 | 2.642 | 3.472 | 9.172 |
| Сернистый газ | $SO_2$ | 1042.7 | 430.64 | 0.284 | 2.421 | 3.521 | 8.526 |

| | | | | | | |
|---|---|---|---|---|---|---|
| Ксенон | $Xe$ | 792.81 | 289.73 | 0.286 | 2.736 | 3.497 | 9.568 |
| Бензол | $C_6H_6$ | 1326.5 | 562.05 | 0.271 | 2.360 | 3.690 | 8.709 |
| Фтор | $F_2$ | 383.62 | 144.41 | 0.288 | 2.656 | 3.472 | 9.224 |
| Сероводород | $H_2S$ | 942.63 | 373.1 | 0.284 | 2.526 | 3.521 | 8.896 |

**Табл. 2.** Результаты статистического анализа

| $f$ | $<f>$ | $\delta f$ | $\delta f/<f>$ | $\Delta f$ | $\Delta f/<f>$ |
|---|---|---|---|---|---|
| $T_B/T_c$ | 2.560 | 0.156 | 0.061 | 0.130 | 0.051 |
| $z_c$ | 0.285 | 0.012 | 0.042 | 0.008 | 0.028 |
| $T_B/T_c/z_c$ | 8.978 | 0.318 | 0.035 | 0.284 | 0.032 |
| $1/z_c$ | 3.513 | 0.142 | 0.040 | 0.103 | 0.029 |

Отметим, что относительная среднеквадратичная ошибка
$\delta((T_B/T_c)/z_c)/<(T_B/T_c)/z_c>= 0.035$ в три раза меньше суммы таких же ошибок
$\delta(T_B/T_c)/<T_B/T_c>+\delta(z_c)/<z_c>= 0.103$, а относительная ошибка
$\Delta((T_B/T_c)/z_c)/<(T_B/T_c)/z_c>= 0.032$ в 2.5 раза меньше суммы относительных ошибок
$\Delta(T_B/T_c)/<T_B/T_c>+\Delta(z_c)/<z_c>= 0.079$.

Для коэффициента корреляции (стр.186 в [10]) величин $T_B/T_c$ и $1/z_c$ имеем

$$\sum_{j=1}^{N}\frac{T_{Bj}/T_{cj}-<T_B/T_c>}{\delta(T_B/T_c)}\cdot\frac{1/z_{cj}-<1/z_c>}{\delta(1/z_c)}\cdot\frac{1}{N}=-0.817\cdot$$

Коэффициент корреляции равен 0 для двух несвязанных (некоррелированных) между собой величин. Если коэффициент корреляции был бы равен -1, то между этими величинами существовала бы прямая пропорциональность (с.187 в [10]). То, что коэффициент корреляции величин $T_B/T_c$ и $1/z_c$ близок к -1 свидетельствует о наличии сильной корреляции (связи) между ними.

Если считать, что величины $T_B/T_c$ и $1/z_c$ независимы, то дисперсия (ошибка) определения $S$ произведения равна (стр.203 в [10])

$$S=\sqrt{<1/z_c>^2\cdot\delta^2(T_B/T_c)+<T_B/T_c>^2\cdot\delta^2(1/z_c)+\delta^2(T_B/T_c)\cdot\delta^2(1/z_c)}=0.658,$$

и поэтому $\frac{S}{<(T_B/T_c)>\cdot<1/z_c>}=0.073\cdot$

Следовательно, произведение этих двух величин должно иметь разброс, равный 7.3%, в то время как в действительности для 21 вещества из табл. 1 разброс в два раза меньше и равен 3.5%. Это также показывает, что эти величины не являются независимыми.

Значения температур Бойля в таблице 1 являются надежными и высокоточными, так как они получены на основе современных стандартных уравнений состояния, описывающих $(p,v,T)$ – данные веществ с высокой точностью [9]. Например, $(p,v,T)$ – данные для аргона [11] стандартное уравнение [12] описывает с точностью 0.03%.

Из вышеизложенного статистического анализа следует, что для всех однокомпонентных веществ из табл. 1 с хорошей точностью имеет место связь (5).

Выполнение связи (5) с хорошей точностью для 21 вещества различного химического состава с различными молекулярными пространственными структурами (одноатомные, молекулярные – линейные и нелинейные) и различными потенциалами взаимодействия между молекулами (сферически симметричный и анизотропный потенциалы) свидетельствует в пользу того, что эта связь является неслучайной.

Несмотря на то, что критическая точка была открыта более чем 140 лет назад, свойства вещества в этой точке окончательно не изучены и по сей день ведутся активные исследования этой точки (смотрите, например, [15-23]).

## 3. Другие связи между термодинамическими параметрами

Как видно из табл. 1 и соотношений (3), для большинства веществ из этой таблицы: метана, этана, пропана, бутана, изобутана, окиси углерода, аргона, криптона, ксенона, азота, кислорода, сернистого газа, фтора и сероводорода имеет место приближенное равенство

$$z_c = \frac{T_{cvdw}}{T_{Bvdw}} = 0.296 \;.$$

Кроме того для этих же веществ имеет место приближенное соотношение
$$\frac{T_B}{T_c} = \frac{1}{z_{cvdw}} = 2.667 \;.$$

Формула Филиппова-Тиммерманса [5] $z_c = \frac{v_{00}}{v_c}$, где $v_{00} = m/\rho_{00}$, $m$ – масса частицы, из которой состоит вещество, $\rho_{00}$ – ортобарическая плотность жидкости, экстраполированная к нулевому значению абсолютной температуры, и соотношение (5) дают $\frac{v_c}{T_c} = 9\frac{v_{00}}{T_B}$. (6)

Из соотношения (5) и равенства $z_c = p_c v_c / k T_c$ получаем $p_c v_c = \frac{k T_B}{9}$. (7)

Наиболее трудным для экспериментального определения критическим параметром является критическая плотность, обратно пропорциональная критическому объему [13]. Это связано с резким увеличением флуктуаций плотности при приближении к критической точке.

Поэтому важно иметь соотношения, позволяющие определить критический объем через относительно легко определяемые характеристики вещества.

К ним относятся полученные нами новые соотношения (6) и (7). Соотношение (6) позволяет определить критический объем $v_c$ через известные критическую температуру и параметры $v_{00}$ и $T_B$, а соотношение (7) позволяет определить критический объем через известные температуру Бойля и критическое давление.

Многие металлы, в том числе металлы и их сплавы, являющиеся конструкционными материалами, а также металлы и их соединения, являющиеся рабочими телами и теплоносителями в энергетических установках (в том числе ядерных), имеют аномально высокие значения критических давления и температуры [14].

При таких высоких значениях этих параметров конструкционные материалы, из которых изготовлено оборудование, использующееся в обычных методах физического исследования свойств веществ, плавятся (высокие температуры), либо оборудование разрушается (высокие давление и температура).

Кроме того при критической плотности может начаться неконтролируемое деление радиоактивных изотопов металлов и их соединений. Поэтому важными как для научных, так и для практических целей являются соотношения, связывающие критические параметры с другими параметрами вещества при низкой температуре и/или низком давлении и/или малой плотности. К ним относятся соотношения (6) и (7).

Параметры $T_B$ в (6)-(7) и $v_{00}$ в (6) относятся к прямой линии единичной сжимаемости $v_{00}/v + T/T_B = 1$ на плоскости $(T, 1/v)$. Эта линия существует в интервале от температуры тройной точки до критической температуры для жидкости и от критической температуры до температуры $T_B$ для вещества в сверхкритическом состоянии [2].

Поэтому параметры $v_{00}$ и $T_B$ могут быть определены как из характеристик вещества в сверхкритическом состоянии (низкое давление и малая плотность) так и из характеристик жидкости вблизи тройной точки (низкие температура и давление).

Кроме того $T_B$ может быть определена из второго вириального коэффициента (низкое давление и малая плотность) [6].

Соотношение (7) связывает важную характеристику критической точки – произведение критического давления и критического объема с температурой Бойля. Температура Бойля зависит только от потенциала взаимодействия между двумя частицами [6].

Следовательно, произведение $p_c v_c$ полностью определяется через параметры потенциала взаимодействия между двумя частицами. Это указывает на важность двойных столкновений вблизи критической точки, и позволяет предположить, что свойства вещества вблизи критической точки определяются двойными столкновениями.

**Выводы**

1. Уравнение состояния Ван-дер-Ваальса дает соотношение между критическими параметрами и температурой Бойля, выполняющееся с достаточной точностью для 21 однокомпонентного вещества, в то время как оно только качественно описывает ($p,v,T$) – свойства и фазовый переход газ-жидкость однокомпонентных веществ. Поэтому рано списывать это уравнение в исторический материал как отслужившее свою роль в физике и химии.

2. Один из критических параметров 21 реального вещества может быть определен через температуру Бойля, что позволяет допустить, что один из критических параметров других веществ, подобных этим, также определяется через температуру Бойля.

# The relation of the parameters of the critical point of liquid-gas transition with the Boyle temperature

Umirzakov I. H.

*Institute of Thermophysics, Novosibirsk, Russia*
*e-mail: tepliza@academ.org*

**Abstract**

It is shown that the ratio of the Boyle temperature to the product of critical temperature and critical compressibility factor is equal to the number 9 with high accuracy for 21 real substances as predicted from Van-der-Waals equation of state. The relation is suggested to find the critical volume via the ratio of the Boyle temperature to the critical pressure. The formula is suggested also to define the critical volume via the critical temperature and the parameters of the linear line of unite compressibility.

***Keywords***: *critical temperature, critical volume, critical pressure, Boyle temperature, Van-der-Waals equation of state, liquid-gas phase transition.*